\documentclass[12pt]{article}

\usepackage{fullpage}
\usepackage{graphicx,amsmath}
\usepackage{amsfonts,amssymb,latexsym,textcomp}
\usepackage{epsfig}
\usepackage{hyperref}
\usepackage{float}
\usepackage{multirow}
\usepackage{rotating}
\usepackage{color}
\usepackage{subfigure}
\usepackage{algorithm}

\begin{document}

\title{SCALABLE INTERNETWORKING \\
Final Technical  Report}
\author{J.J. Garcia-Luna-Aceves and Anujan Varma \\
Computer Engineering Department \\
University of California \\
Santa Cruz, CA 95064} 
\date{}

\maketitle

This document describes the work completed at the University of
California, Santa Cruz under the project ``Scalable Internetworking''
sponsored by ARPA under Contract No. F19628-93-C-0175. This report
covers work performed from 1 April 1993 to 31 December 1995.

Results on routing and multicasting for large-scale internets are
summarized. The technical material discussed assumes familiarity with
the content of our proposal and previous quarterly reports submitted
in this project.

\newpage

\section{INTRODUCTION}

Today's internetworking technology is challenged by growth, the
provision of multiple types of services at varying speeds, the support
of collaborative environments that require real-time multipoint
communications, the control of resources by multiple administrative
authorities, and the interoperation of ATM-based internets with
connectionless networks and internets.  In the long term, providing
adequate internet routing that accommodates its size and diversity of
service requirements cannot be met with traditional link-state or
distance vector routing algorithms, and a strategy is needed for the
interconnection of large numbers of connectionless networks through
ATM.

Funded by the Advanced Research Projects Agency (ARPA), the University
of California, Santa Cruz (UCSC) carried out research on {\em scalable
internetworking} from 1 April 1993 to 31 December 1995.  The original
goals of this project were to advance the state of the art in
internetworking technology by
\begin{itemize}
\item
Developing and validating the first new type of routing algorithm
(which we call link-vector algorithms or LVA) since the development of
the ARPANET's routing protocols, which introduced the first
distance-vector and link-state algorithms for computer networks.  LVA
implements the selective diffusion of link states based on distributed
computation of preferred paths that can take into account service
types, policy restrictions, and the characteristics of transmission
media.
\item
Unifying previous work on the correctness of distance-vector and
link-state algorithms as a problem of distributed termination
detection.
\item
Developing more efficient mechanisms for termination detection
distance-vector and link-state algorithms, which can be used to
improve the performance of existing interdomain and intradomain
routing protocols.
\item
Developing new distributed approaches to multicasting capable of
supporting multiple types of service within the same multipoint
session.
\item
Applying the new algorithms summarized above, plus novel approaches to
the mapping of names into addresses and routes, to the interconnection
of connectionless LANs and MANs through ATM-based internets, and to
fully-integrated ATM-based internets.
\end{itemize}

In the course of our research, we developed new ideas on routing and
multicasting that we had not predicted as part of our original goals.
The results of our research have been made available in the open
literature as part of the following 20 publications:

\begin{enumerate}

\item F. Bauer and A. Varma,
``Degree-Constrained Multicasting in Point-to-Point Networks,''
{\em Proceedings of IEEE INFOCOM '95}, April 1995.

\item F. Bauer and A. Varma,
``Distributed Algorithms for Multicast Path Setup in Data Networks''
{\em IEEE/ACM Transactions on Networking}, to appear.

\item F. Bauer and A. Varma,
``Distributed Algorithms for Multicast Path Setup in Data Networks,''
{\em Proceedings of IEEE GLOBECOM~'95}, November 1995.

\item F. Bauer and A. Varma,
``ARIES: A Rearrangeable Inexpensive Edge-based On-line Steiner Algorithm,''
{\em Proceedings of IEEE INFOCOM~'96}, March 1996, to appear.

\item
J. Behrens and J.J. Garcia-Luna-Aceves, ``Distributed, Scalable
Routing Based on Link-State Vectors,'' {\em Proc. ACM SIGCOMM 94},
London, U.K., August 1994.

\item
J.J. Garcia-Luna-Aceves and S. Murthy, A Path-Finding Algorithm for
Loop-Free Routing,'' accepted for publication in {\em IEEE/ACM
Transactions on Networking}, 1996.

\item
J.J. Garcia-Luna-Aceves and J. Behrens, ``Distributed, Scalable
Routing based on Vectors of Link States,'' {\em IEEE Journal on
Selected Areas in Communications}, Vol 13, No. 8, October 1995.

\item 
J.J. Garcia-Luna-Aceves and S. Murthy, ``A Loop-Free Path-Finding
Algorithm: Specification, Verification and Complexity,'' {\em
Proc. IEEE INFOCOM 95}, Boston, MA, April 1995.

\item
J.J. Garcia-Luna-Aceves and Y. Zhang, ``Reliable Broadcasting in
Dynamic Networks,'' {\em Proc. IEEE ICC 96}, Dallas, Texas, June
23-27, 1996.

\item L.~Kalampoukas, A.~Varma, and K.~K.~Ramakrishnan, ``An
Efficient Rate Allocation Algorithm for Packet-Switched Networks
Providing Max-Min Fairness,'' {\em Proc. IFIP Conference on
High-Performance Networks (HPN 95)}, September 1995.

\item L. Kalampoukas, A. Varma and K. K. Ramakrishnan, "Dynamics of an
Explicit Rate Allocation Algorithm for Available Bit-Rate (ABR)
Service in ATM Networks", {\em Proc. IFIP Conference on Broadband
Networks}, Montreal, Canada, April 1996, to appear.

\item L. Kalampoukas and A. Varma, "Analysis of Source Policy in
Rate-Controlled ATM Networks," {\em Proc. ICC '96}, to appear.

\item
S. Murthy and J.J. Garcia-Luna-Aceves, ``A Routing Protocol for
Wireless Networks,'' accepted for publication in {\em ACM NOMAD
Journal}, 1996.

\item
S. Murthy and J.J. Garcia-Luna-Aceves, ``Congestion-Oriented
Shortest-Multipath Routing,'' {\em Proc. IEEE INFOCOM 96}, San
Francisco, California, March 1996.

\item
S. Murthy and J.J. Garcia-Luna-Aceves, ``A Loop-Free Path-Finding
Algorithm: Analysis of Dynamics,'' {\em Proc. IEEE Globecom 95},
Singapore, November 13--17, 1995.

\item
S. Murthy and J.J. Garcia-Luna-Aceves, ``A Routing Protocol for
Packet-Radio Networks,'' {\em Proc. First ACM International Conference
on Mobile Computing and Networking}, Berkeley, California, November
13--15, 1995.

\item
S. Murthy and J.J. Garcia-Luna-Aceves, ``A More Efficient Path-Finding
Algorithm,'' {\em Proc. Twenty-Eighth Annual Asilomar Conference on
Signals, Systems and Computers}, Pacific Grove, California, October
31- November 2, 1994. 

\item
S. Murthy and J.J. Garcia-Luna-Aceves, ``A Loop-Free Algorithm Based
on Predecessor Information,'' {\em Proc. Third International
Conference on Computer Communications and Networks---ICCCN 94}, San
Francisco, California, September 11-14, 1994.

\item
M. Parsa and J.J. Garcia-Luna-Aceves, ``Scalable Internet Multicast
Routing,'' {\em Proc. ICCCN 95}, Las Vegas, Nevada, September 20--23,
1995.

\item
 Q.  Zhu, M. Parsa, and J.J. Garcia-Luna-Aceves, A Source-Based
Algorithm for Delay-Constrained Minimum-Cost Multicasting,'' {\em
Proc. IEEE INFOCOM 95}, Boston, MA, April 1995.

\end{enumerate}

All these publications are available online at the following WWW page:
\\ http://www.cse.ucsc.edu/research/ccrg/projects.internet.html.  A
few of our most recent papers are included at the end of this report.

Our results on routing and multicasting are already being applied in
other ARPA projects. Specifically, as part of the project Wireless
Internet Gateways (WINGS), we are using path-finding algorithms to
implement routing protocols for wireless networks.  

The rest of this report summarizes the main results of our
research. Each section points to the associated publications in the
previous list, where the details of our research are presented.

\section{ROUTING ALGORITHMS} 

\subsection{Link Vector Algorithms (LVA)}

Our results in this area are presented in Publications 5 and 7.

We developed a new type of routing algorithm, which we call link
vector algorithm (LVA).  The basic idea of LVAs consists of asking
each router to report to its neighbors the characteristics of each of
the links it uses to reach a destination through one or more preferred
paths, and to report to its neighbors which links it has erased from
its preferred paths.  Using this information, each router constructs a
source graph consisting of all the links it uses in the preferred
paths to each destination.  LVAs ensure that the link-state
information maintained at each router corresponds to the link
constituency of the preferred paths used by routers in the network or
internet.  Each router runs a local algorithm or multiple algorithms
on its topology table to compute its source graph with the preferred
paths to each destination.  Such algorithm can be any type of
algorithm (e.g., shortest path, maximum-capacity path, policy path)
and the only requirements for correct operation are for all routers to
use the same algorithm to compute the same type of preferred paths,
and that routers report all the links used in all preferred paths
obtained.  Aggregation of information can take place adapting the
area-based routing techniques proposed for DVAs in the past.

Because LVAs propagate link-state information by diffusing link states
selectively based on the distributed computation of preferred paths,
LVA reduce the communication overhead incurred in traditional LSAs
based on flooding of link states.  Because LVAs exchange routing
information that is related to link (and even node) characteristics,
rather than path characteristics, LVAs reduce the combinatorial
explosion incurred with any type of DVA for routing under multiple
constraints.  The simulation results obtained for a particular LVA
shows that it has better performance than an ideal link-state
algorithm based on flooding and the distributed Bellman-Ford
algorithm.

An important contribution of this research was to show that LVAs are
correct under different types of routing, assuming that a correct
mechanism is used for routers to ascertain which updates are recent or
outdated.

LVAs open up a large number of interesting possibilities for internet
routing protocols. To name a few, LVAs can be the basis for effective
routing protocols based on link-state information for packet radio
networks.  LVAs can be used to develop more efficient intra-domain
routing protocols that are based on link-state information but require
no backbones (which eliminates the difficult network-management
problems associated with such protocols as OSPF and ISO IS-IS) and can
take advantage of aggregation schemes developed for distance-vector
algorithms. Finally, LVAs make path-vector algorithms (used in BGP and
IDRP, for example) obsolete. LVAs are applicable to the emerging
Nimrod architecture and protocols being developed under ARPA
sponsorship.

\subsection{Path-Finding Algorithms}

Our results in this area are presented in Publications
6, 8, 13, and 15--18.

Recently, distributed shortest-path algorithms that utilize
information regarding the length and second-to-last hop (or
predecessor) of the shortest path to each destination have been
proposed to eliminate the counting-to-infinity problem of the
distributed Bellman-Ford algorithm (DBF), which is used in a number of
today's routing protocols.  We call these type of algorithms
path-finding algorithms.  Although these algorithms provide a marked
improvement over DBF, they do not eliminate the possibility of
temporary loops.  The loop-free algorithms reported to date rely on
mechanisms that require routes either to synchronize along multiple
hops, or exchange path information that can include all the nodes in
the path from source to destination.

We developed the first known path-finding algorithm that is loop-free
at every instant.  We call this path-finding algorithm the loop-free
path-finding algorithm (LPA).  According to LPA, update messages are
sent only to neighboring nodes.  Like previous path-finding
algorithms, LPA eliminates the counting-to-infinity problem of DBF
using the predecessor information.  In addition, LPA eliminates all
temporary loops by implementing an interneighbor coordination
mechanism with which potential temporary loops are blocked before
routers can forward data through them.  To block a potential temporary
loop, a node sends a query to all its neighbors reporting an infinite
distance to a destination, before changing its routing table; the node
is free to choose a new successor only when it receives the replies
from its neighbors.  To reduce the communication overhead incurred
with interneighbor coordination, nodes use a {\em feasibility
condition} to limit the number of times when they have to send queries
to their neighbors.  In contrast to many prior loop-free routing
algorithms, queries propagate only one hop in LPA.  Furthermore,
updates and routing-table entries in LPA require a single node
identifier as path information, rather than a variable number of node
identifiers as in prior algorithms.

We compared LPA's performance against the performance of the most
efficient loop-free algorithm previously known (DUAL) and an ideal
link-state algorithm (ILS). The simulation considered the dynamics of
these three algorithms after a single event (link or node failure,
link or node addition, and link-cost change) and multiple link-cost
changes. Our simulation results clearly indicate that LPA is far more
scalable than DUAL and ILS. The results obtained for multiple
link-cost changes show that LPA is the first distance-vector algorithm
to require less overhead traffic than any link-state algorithm based
on flooding.

\section{CONGESTION-ORIENTED SHORTEST MULTIPATH ROUTING}

Our results in this are are presented in Publication 14.

We have demonstrated that a connectionless routing architecture can
provide performance guarantees.  We used architectural elements
similar to those used in a connection-oriented architecture to allow
the network to enforce performance guarantees in the delivery of those
datagrams that are accepted into the network.  

In our new framework for dynamic multipath routing in connectionless
internets, packets are individually routed towards their destinations
on a hop by hop basis. A packet intended for a given destination is
allowed to enter the network if and only if there is at least one path
of routers with enough resources to ensure its delivery within a
finite time. In contrast to existing connectionless routing schemes,
once a packet is accepted into the network, it is delivered to its
destination, unless resource failures prevent it. Each router reserves
buffer space for each destination, rather than for each
source-destination session as it is customary in a connection-oriented
architecture, and forwards a received packet along one of multiple
loop-free paths towards the destination. The buffer space and
available paths for each destination are updated to adapt to
congestion and topology changes.

Parekh and Gallager have analyzed worst-case session delay in a
connection-oriented network architecture~\cite{parekh94}.  We have
obtained a similar upper bound on the steady-state delay experienced
by a datagram accepted into the network using our new routing
framework; this upper bound is given by the following equation:

\begin{equation} \label{Eqnyyy}
D^{i*}_j(t) < \Delta^i_j(t) [1 + Q^{i*}_j(t)] + MAD^i_j(t) 
\end{equation}

\noindent
where $D^{i*}_j{t}$ is the longest delay that can be experienced by a
packet created or forwarded by node $i$ to destination $j$ at time
$t$, $Q^{i*}_j(t)$ is the worst-case delay of transmitting packets
already in queue for destination $j$, and $MAD^i_j(t)$ is the maximum
delay allowed by the routing protocol for any neighbor of node $i$ to
be considered part of a loop-free path to destination $j$.  The first
term of Eq.~\ref{Eqnyyy} corresponds to the delay incurred by sending
all backlogged packets at time $t$ to a neighbor with the longest link
propagation delay (which is denoted by $\Delta^i_j(t)$). The second
term corresponds to the maximum delay incurred by any neighbor
receiving the backlog packets; because any such neighbor must be on
the ``shortest multipath'' from $i$ to destination $j$ (i.e., a set of
loop-free paths with a maximum delay smaller than $MAD^i_j(t)$), that
delay can be at most equal to $MAD^i_j(t)$.

The above bound is the {\em first} known upper bound for packet delays
in a connectionless routing architecture, and is possible because
datagrams are accepted only if routers have enough credits to ensure
their delivery, and datagrams are delivered along loop-free paths. In
contrast, in traditional datagram routing architectures, any datagram
presented to a router is sent towards the destination, and the paths
taken by such datagrams can have loops; therefore, it is not possible
to ensure a finite delay for the entry router or any relay router
servicing a datagram.

\section{INTERNET MULTICASTING}

\subsection{Delay-Constrained Minimum-Cost Multicasting}

Our results in this area are presented in Publication 20.

Different optimization goals can be used in multicast routing
algorithms to determine what constitutes a good tree.  One such goal
is providing minimum delay along the tree, which is important for such
multimedia applications as real-time conferencing.  Another
optimization goal is making use of the network resources as
efficiently as possible; two interesting variants of this objective
are
\begin{itemize}
\item Minimizing the total bandwidth utilization of links; we call this
objective {\em utilization driven}, because it minimizes the total
bandwidth utilization cost of links for a data stream sent from the
source to destinations.
\item Distributing bandwidth utilization of sessions
among links in the tree
in order to minimize congestion along links; we call this objective
{\em congestion driven}, because it minimizes the maximal link cost
requirement along the transferring paths.
\end{itemize}

Previous optimization techniques for multicast routing algorithms have
considered the above two optimization objectives, but have treated
them as distinct problems.  Dijkstra's shortest path algorithm
\cite{Dijkstra59} can be used to generate the shortest paths from the
source to destination nodes; this provides the optimal solution for
delay optimization. Multicast routing algorithms that perform cost
optimization have been based on minimum Steiner tree which is known to
be NP-complete problem \cite{Garey79}. Some heuristics for the Steiner
tree problem \cite{Kou81,Rayward-Smith83,Takahashi80} have
been developed that take polynomial time and produce near optimum
results.  In Kou, Markowsky and Berman's (KMB) algorithm \cite{Kou81},
a network is abstracted to a complete graph consisting of edges that
represent the shortest paths among source node and destination
nodes. The KMB algorithm applies Prim's algorithm \cite{Prim57} to
construct a {\em minimum spanning tree} in the complete graph, and the
Steiner tree of the original network is obtained by achieving the
shortest paths represented by edges in the minimum spanning tree.
Waxman \cite{Waxman88} examined the dynamic update of the tree if
destination nodes join or leave the tree occasionally.  Shacham and
Meditch \cite{Shacham93} investigated the maximum flow distribution of
multiple streams along a multicast tree; Optimum assignment is based
on link capacity and destination requests, but the request is not
delay-bounded and a dynamic programming algorithm is devised that may
take exponential time complexity.

Bharath-Kumar and Jaffe discuss optimization on both cost and delay
\cite{Kadaba83}. However, they assume that cost and delay functions
are identical.  Kompella, Pasquale and Polyzos \cite{Kompella93}
propose two heuristics (which we call the KPP algorithm) that address
delay-bounded Steiner trees; the KPP algorithm extends the KMB
algorithm by taking into account an integer-valued delay bound in the
construction of shortest paths.

We developed a new algorithm for multicast tree construction that we
call {\em Bounded Shortest Multicast Algorithm} (BSMA).  BSMA is {\em
source based}, which means the source of the multicast is assumed to
know all the information needed to construct the multicast
tree. Although the model of delay-bounded minimum Steiner tree is
similar to the one used in the KPP algorithm, we generalize the
formulation of the problem to more realistic network settings; the
major contributions of the new algorithm, which we call {\em Bounded
Shortest Multicast Algorithm} (BSMA)
can be summarized as follows:
\begin {itemize}
\item  It considers the congestion-driven and 
utilization-driven variants of the cost function.
\item  
It specifies variable delay bounds reserved for destination nodes,
that is, different destinations have variable delay bounds;
furthermore, a delay bound can take arbitrary real values.
\item 
Instead of using a single pass as in previous algorithms, it
iteratively (in multiple passes) minimizes the cost function of the
tree.
\end{itemize}

Instead of the single-pass tree construction approach used in most
previous heuristics, BSMA is based on a feasible search optimization
method, starting from the minimum-delay tree, and monotonically
decreases the cost by iterative refinement of the delay-bounded tree.
The optimality of the costs of the delay-bounded trees obtained with
BSMA were analyzed by simulation; our results show that, depending on
how tight the delay bounds are, the costs of the multicast trees
obtained with BSMA are very close to the costs of the trees obtained
by a well-known, provably near-optimal minimum Steiner tree heuristic,
with the costs within the factor $2(1-\frac{1}{|S|})$ of the optimal
Steiner tree for $|S|$ connected nodes.

\subsection{Multicast Internet Protocol (MIP)}

Our results in this area are presented in Publication 19 and the
following paper:
\begin{itemize}
\item
M. Parsa, and J.J.~Garcia-Luna-Aceves, ``A Protocol for Scalable
Loop-Free Multicast Routing,'' submitted to {\em IEEE Journal on
Selected Areas in Communications}, 1996.
\end{itemize}

Multicasting is supported in local area networks (LANs), using
hardware technologies.  Recently, multicasting has been extended to
internetworks by Deering \cite{Deering90}.  Based on Deering's work
and built into the TCP/IP protocol suite, the internet group message
protocol (IGMP) is used to disseminate multicast membership
information to multicast routers.  Deering's method permits routers to
figure out dynamically how to forward messages.  A delivery tree is
constructed on-demand and is data-driven.  The tree in the existing IP
architecture is the reverse shortest-path tree and shortest-path tree
from the source to the group for distance-vector (i.e., DVMRP
\cite{Waitzman}) and link-state routing (i.e., MOSPF \cite{Moy94}),
respectively.  For example, the Multicast Backbone (MBone) in today's
Internet consists of a set of routers running DVMRP.  However, there
are several shortcomings with the existing IP multicast architecture,
i.e., DVMRP and MOSPF.  First, all routers in the Internet have to
periodically generate and process control message for $every$
multicast group, regardless of whether or not they belong to the
multicast tree of the group.  Thus, routers not on the multicast tree
incur memory and processing overhead to construct and maintain the
tree for the lifetime of the group.  The packets that are periodically
flooded throughout the Internet but that do not lead to any receivers
or sources consume bandwidth.  In DVMRP, it is the data packets that
are periodically flooded when the state information for a multicast
tree times out.  In MOSPF, it is the link-state packets, containing
the state information for group membership, that are periodically
flooded.  Second, the multicast routing information in each router is
stored for each source sending to a group.  If there are $S$ sources
and $G$ groups, the multicast protocols scale as $\Theta(SG)$.
Finally, the IP multicast protocols, being extensions of unicast
routings, are tightly coupled to the underlying unicast routing
algorithm.  This complicates inter-domain multicasting if the domains
involved use different unicast routing.  The unicast routing is also
made more complicated by incorporating the multicast-related
requirements.

To overcome the above shortcomings, two protocols have been recently
proposed: the core-based tree (CBT) architecture \cite{Ballardie93}
and the protocol independent multicast (PIM) architecture
\cite{Deering94}.  Although both approaches constitute a substantial
improvement over the current multicast architecture, each protocol has
its own limitations.  A main limitation of CBT is that it constructs
only a single tree per group and thus provides longer end-to-end
delays than would be obtained along a shortest-path tree.  A main
limitation of PIM is that the periodic control messages, i.e., its
soft-state mechanism, incur overhead even in an stable internet.  We
have shown that both protocols suffer from temporary loops resulting
from the use of inconsistent unicast routing information, and neither
protocol has been verified to provide correct multicast routing trees
after network changes.

We developed a new multicast routing protocol called {\it Multicast
Internet Protocol} (MIP), which solves the shortcomings of the
previous approaches to multicast routing.  MIP offers a simple and
flexible approach to the construction of both group-shared and
shortest-path multicast trees.  The shortest-path trees in MIP can be
relaxed to cost-bounded trees, making it possible to trade off
optimality of the tree with the control message overhead of
maintaining a shortest-path tree.  MIP accommodates sender-initiated
and receiver-initiated multicast tree construction, which makes MIP
flexible to use for a wide range of applications with different
characteristics, group dynamics and sizes.  Although MIP is
independent of the underlying unicast routing, MIP never creates loops
in a multicast tree.  Since applications for group communication tend
to be bandwidth intensive, for example, from using images and video,
loop-freedom is a specially important consideration in multicasting.
Instead of using the idea of ``soft state'' to maintain multicast
routing information, MIP uses diffusing computations to update and
disseminate multicast routing information and to insure loop-freedom.
This last feature of MIP has a number of scaling properties: under
stable network conditions, MIP has no control message overhead to
maintain multicast routing information; MIP responds to network events
as fast as routers can propagate update information, rather than
waiting for timers to expire before propagating changes; finally,
because no loops can occur, routers obtain correct multicast routing
or stop forwarding data to a portion of a multicast tree as fast as
update information can propagate along a desired multicast tree.

\section{RELIABLE BROADCASTING}

Our results in this area are presented in Publication 9.

Network broadcasting consists of delivering copies of the same message
to all the nodes in a network. The information that is to be
broadcasted can be varied, for example, a command, the information of
users' location, the information about users' connectivity, or a
network topology change due to link failures and repairs.  The focus
of this paper is on the reliable broadcasting of information in
wireless networks with dynamic topologies.

Reliable broadcasting is not trivial when the topology of the network
can change due to failure or mobility, whether or not the source of
information has a local copy of the topology.  A reliable broadcast
protocol must ensure that the information from the source node reaches
all the nodes that are connected to the source node in the network
within a finite time, and that the source node is notified about that.
A reliable broadcast protocol for dynamic networks should
\begin{itemize}
\item
Operate independently of the routing information available at nodes,
which may be inaccurate, and without requiring the source to know
which nodes have a physical path to it
\item
Incur low latency and as few messages as possible in order to broadcast a
message from any given source.
\end{itemize}

Surprisingly, little work exists on reliable broadcast protocols in
dynamic networks; most broadcast protocols proposed in the past (e.g.,
\cite{SEGA-AWER-83}) \cite{SEGA-83} are based on what Segall calls the
propagation of information (PI) protocol and the propagation of
information with feedback protocol (PIF) \cite{SEGA-83}.  PI is basic
flooding; it starts by the source node sending a message containing
the information to all its neighbors, and each other node in the
network sends a similar message to its own neighbors when it receives
the first such message.  PIF is based on PI, with the source node
being informed when the information has reached all connected nodes in
the network. The difference between PIF and PI is that, when a node
$i$ receives the first message from a neighbor node $p_i$, node $i$
labels that neighbor node as its successor. When a node $i$ other than
the source has received the source's message from all neighbors, it
sends the source's message back to its successor $p_i$.  PIF
terminates when the source node receives its own message from all its
neighbors. 

All reliable broadcast protocols based on flooding that have been
proposed for dynamic topologies in the past (e.g.,
\cite{SEGA-AWER-83}) are based on the routing protocol by Merlin and
Segall \cite{MERL-79}.  These protocols proceed in cycles triggered
and terminating at the source of the message. A cycle consists of two
phases. First, the message propagates one additional hop away from the
source, then acknowledgments to the message propagate towards the
source. The source starts by sending a message that is acknowledged by
all its neighbors. When the source receives the acknowledgments from
its neighbors, it then resends the message asking the neighbors to
propagate the message one more hop to their own neighbors. The
neighbors forward the message to their neighbors and send their
acknowledgments back to the source when they receive the
acknowledgments from all their own neighbors, and so forth. This
scheme incurs too much communication overhead to be attractive for a
wireless network; furthermore, an implicit assumption of this approach
is that a node can have a fairly stable successor to the source, which
does not apply in a network with mobile nodes.

We developed a new reliable broadcast protocol that is correct for
both static and dynamic networks. We call this protocol the reliable
broadcast protocol (RBP).

RBP ensures that every node connected to the source node receives the
information from the source node at least once, and that the source
node is positively informed that the information reaches all the
connected nodes in the network within a finite time.  RBP works in a
similar way to PIF for the case of a static network.  However, in
contrast to prior approaches of reliable broadcasting in dynamic
networks, our proposed protocol requires the source to send each
broadcast message only once, and diffusing computations \cite{DIJK-80}
are used to eliminate the need for the source node to control the
propagation of information in multiple rounds.  An important
contribution of this paper is the adaptation of Dijkstra and
Scholten's basic scheme to dynamic topologies. Instead of defining a
single successor for each node in a directed acyclic graph (DAG), RBP
permits each node to define a successor set formed with all those
neighbors who transmit the source's message.  To our knowledge, RBP is
the first protocol for reliable broadcasting that is not based on PIF
or pre-established trees.

\section{MULTICASTING IN ATM NETWORKS}

Our results in this area are presented in Publications 1--4.

We developed and evaluated a number of heuristic algorithms for
finding efficient multicast trees in the presence of constraints on
the copying ability of the individual switch nodes in an ATM network.
We studied bot centralized and distributed heuristics for solution of
the degree-constrained Steiner problem.  Two distinct approaches were
considered for the design of distributed DCPS heuristics.  The first
approach involves design of distributed versions of centralized DCSP
algorithms.  We introduced distributed versions of two DCSP
heuristics, SPH and K-SPH.  The second approach is to modify the
solution obtained from an unconstrained heuristic to satisfy the
degree constraints using a distributed post-processing algorithm.  We
compared the algorithms by simulation based on three criteria: quality
of solution, convergence time, and the number of unsolved networks.
Our results show that each of the heuristics generated
degree-constrained multicast trees within 10\% of the best solution
found.  Surprisingly few test networks were unsolvable.  The
distributed post-processing heuristic presented is of particular
interest since it may be used with any Steiner heuristic.  When paired
with a good unconstrained distributed Steiner heuristic, it gave away
little quality of solution while converging rapidly.

We introduced and evaluated a new heurtistic, ARIES, for updating
multicast trees dynamically in large point-to-point networks.  The
algorithm is based on monitoring the accumulated damage to the
multicast tree within local regions of the tree as nodes are added and
deleted, and triggering a rearrangement when the number of changes
within a connected subtree crosses a set threshold.  We derived an
analytical upper-bound on the competitiveness of the algorithm.  We
have also obtained simulation results to compare the average-case
performance of the algorithm with two other known algorithms for the
dynamic multicast problem, GREEDY and EBA (Edge-Bounded Algorithm).
Our results show that ARIES provides the best balance among
competitiveness, computational effort, and changes in the multicast
tree after each update.

\section{CONGESTION CONTROL IN ATM NETWORKS}

Our results in this area are presented in Publications 10--12.

Several congestion-control approaches for support of
Available-Bit-Rate (ABR) traffic in ATM networks have been proposed
recently.  These include packet discard schemes, link-level flow
control, and rate control with explicit congestion notification, and
explicit rate setting.  The ATM Forum Traffic Management Committee has
been working on a rate-based approach with explicit rate setting as
the basis of its congestion-control standards for supporting ABR
service.

The rate-based approach requires an algorithm for fair allocation of
bandwidth among the connections sharing a common output link of a
switch.  We designed such an algorithm for rate allocation within the
individual switches of an ATM network implementing a rate-based
congestion control algorithm for best-effort traffic.  The algorithm
performs an allocation in $\Theta (1)$ time, allowing it to be applied
to ATM switches carrying a large number of virtual channels.  When the
total available capacity or the requests of the individual connections
change, the algorithm is shown to converge to the max-min allocation.
The worst-case convergence time is $O(M(2D+\frac{1}{R}))$, where $D$
is the round-trip delay, $M$ is the number of distinct values in the
max-min allocation, and $R$ is the rate at which resource-management
(RM) cells are transmitted by each source.  A patent application on
the rate allocation method was filed during last summer.

Since then, we have studied the dynamics of the rate allocation
algorithm in various network configurations, and demonstrated that the
allocation algorithm is fair and maintains network efficiency in a
variety of network configurations.  We also studied the behavior of
TCP/IP sources using ABR service in a network of switches employing
the rate allocation algorithm; the results show substantial
improvements in fairness and efficiency in comparison with the
performance of TCP on an underlying datagram-based network.  Even in a
dynamic network configuration with frequent opening and closing of ABR
connections, TCP connections were able to make sustained progress with
a switch buffer size as small as 1 Kbyte, providing an average link
utilization of approximately 60\% of the attainable maximum.  We also
studied the performance of ABR traffic in the presence of
higher-priority variable bit-rate (VBR) video traffic and showed that
the overall system achieves high utilization with modest queue sizes
in the switches, and the ABR flows adapt to the available rate in a
fairly short interval.  These results are summarized in Reference [2].

We also analyzed the behavior of the source policy in the rate-based
congestion control algorithm being developed by the ATM Forum and
derived approximate analytical closed-form expressions to describe the
rate-increase process.  These approximations are used to analyze the
impact of the source algorithm on the TCP slow-start process operating
over a rate-controlled ATM network where rate allocation is performed
explicitly in the switches.  The results show that the increase in TCP
congestion window ramp-up time is noticeable when the round-trip delay
is small.  When an idle-detection policy is enabled at the source, the
slow-start process is significantly prolonged, and results in slower
recovery upon a cell loss.  The results are verified by simulation.
Reference [3] contains a summary of this work.

 \end{document}